# Seniors' acceptance of virtual humanoid agents

Anna Esposito[1], Terry Amorese[1], Marialucia Cuciniello[1], Antonietta M. Esposito[2], Alda Troncone[1], Maria Inés Torres[3], Stephan Schlögl[4], Gennaro Cordasco[1]

**Abstract**  This paper reports on a study conducted as part of the EU EMPATHIC project, whose goal is to develop an empathic virtual coach capable of enhancing seniors' well-being, focusing on user requirements and expectations with respect to participants' age and technology experiences (i.e. participants' familiarity with technological devices such as smartphones, laptops, and tablets). The data shows that seniors' favorite technological device is the smartphone, and this device was also the one that scored the highest in terms of easiness to use. We found statistically significant differences on the preferences expressed by seniors toward the gender of the agents. Seniors (independently from their gender) prefer to interact with female humanoid agents on both the pragmatic and hedonic dimensions of an interactive system and are more in favor to commit themselves in a long-lasting interaction with them.  In addition, we found statistically significant effects of the seniors' technology savviness on the hedonic qualities of the proposed interactive systems. Seniors with technological experience felt less motivated and judged the proposed agents less captivating, exciting, and appealing.

## 1 Introduction

Aging engenders several health disorders among which are poor vision, memory loss, fine motor skill impairments and cognitive decline. These impairments provoke social isolation making elders less inclined to preserve their relationship with friends and relatives. In turn, social isolation affects mental well-being and leads to psychological and depressive disorders (Theng & Paye, 2011). Statistics have shown that in

[1] Anna Esposito, Terry Amorese, Marialucia Cuciniello, Alda Troncone, and Gennaro Cordasco are with the Università degli Studi della Campania "Luigi Vanvitelli", Department of Psychology, and IIASS, Italy, emails: iiass.annaesp@tin.it, anna.esposito@unicampania.it

[2] Antonietta M. Esposito is with the Istituto Nazionale di Geofisica e Vulcanologia, Sez. di Napoli Osservatorio Vesuviano, Italy,

[3] Maria Inés Torres is with the Universidad del País Vasco UPV/EHU, Speech Interactive Research Group, Bilbao, Spain

[4] Stephan Schlögl is with the MCI Management Center Innsbruck, Dept. Mgmt., Communication & IT, Innsbruck, Austria

Europe approx. 1.2 million senior citizens are suffering from Parkinson's disease (www.parkinsons.org.uk/content/about-parkinsons), 15% of adults aged 60+ years from a mental disorder (www.who.int/mediacentre/factsheets/fs381/en/), and one-in-six people from anxiety and depression with limited access to therapeutic interventions (Munoz et al, 2010). In addition, it is estimated that 47.5 million people worldwide are living with dementia and that this number is likely to increase in the years to come (http://www.who.int/mediacentre/factsheets/fs381/en/). Consequently, considerable burdens are placed on national health care institutions in terms of medical, social, and care costs associated to the assistance of such people (McDaid et al. 2005). Complex autonomous computer interfaces in the form of embodied conversational agents have been proposed as a solution to these problems (see Hawley et al, 2013 for communication disorders; Parker & Hawley, 2013 for ageing, and Prescott et al, 2013 for companionship) because they can provide an automated on-demand health assistance reducing the abovementioned costs, and lighten human caregivers' workload. Such virtual agents, depending on the user's needs would serve as guides, assistants, information presenters, companions, or simply as a reminder for taking medications. Furthermore, it has been shown that a virtual agent can positively influence elderly people's wellbeing acting on psychological aspects like motivation, self-determination, mood, self-efficacy, and copying capabilities (Gallego-Perez et al., 2013). Nevertheless, the current developers' efforts in implementing such systems are unsuitable to the users' demand because of a lack of attention to their requirements and expectations, as well as, a lack in understanding how the interaction with such complex autonomous ICT (Information Communication Technologies) interfaces affects/enhances human reactions/actions, social perception and meaning-making practices in long-term relationships.

One aspect to focus on with respect to human-agent interaction is the level of user acceptance. This concept was introduced by Davis (1989) in an attempt to explain what leads people to the acceptance of technological devices. To this end, acceptance was defined as: "*A process affected by two main factors:*

*1.   Perceived Usefulness, defined as the degree to which an individual believes that using a particular* [technological device] *would enhance his/her performances*

*2.   Perceived Ease of Use, defined as the degree to which an individual believes that using a particular* [device] *would be free of effort"* (Davis 1989, p. 321)

[… and therefore, simplify the understandings and fulfilments of habitual or unfamiliar tasks]. For aged people, the acceptance of new technologies is an adaptive negotiation between the improvements (quality of life, usefulness, enactment) provided by the offered resource and the struggles required (in terms of the costs, cognitive loads, and environmental changes) to allocate such in their personal environment (Giuliani et al. 2005). In summary, the elders' acceptance of a new technology is entwined to their ability to control it, its easiness to use, as well as, its practical advantages. Beside these practical considerations, the concept of the agent's appearance is a strong factor in fa-voring elders' acceptance of assistive technologies. The user's preference toward the agent's physical and social features – such as face, voice, hair, age, gender, eyes, dressing mode, attractiveness, personality – plays a role. As an example, a recent study conducted with 45 healthy elders, aged 65+ showed that they are able (without being informed) to sense agents' personalities and prefer joyful and practical personalities on both the pragmatic and hedonic dimensions (see section 2.3 for a description of the pragmatic and hedonic dimensions of an interactive system) of the interactive system (Esposito et al. 2018). To our knowledge, there are no systematic investigations devoted to assess the role of these agents' features taking particularly into account seniors' preferences. Gong and Nass (2007) showed that the pairing of a human face with a humanoid voice or a humanoid face with a human voice led to distrustful user reactions toward the agents. Sträfling et al. (2010) showed that students' learning motivations did not change no matter whether they are interacting with a humanoid or animal shaped virtual tutor, or a speech only based interactive system. Ring et al. (2014) showed that humanoid agents with cartoon like or toon shaded semblances are considered more friendly and likable than more realistic humanoid agents. Gender and race, as well as, agent's attractiveness have also been found to impact on users' willingness to interact and consequently influence agents' purpose effectiveness (Guadagno et al. 2007). It has been shown that ICT interfaces with a human face improve employers' productivity (Kong, 2013) and agents with human-like faces trigger in users more positive reactions than agents with animal or iconic faces, independently from the agent's gender (Oh et al. 2016). However, it must be mentioned that participants in these studies, were either students or age was not a variable accounted for. Seniors have been involved in a very limited number of studies. When elders are involved, it has been shown that they clearly enjoy interacting with a speaking synthetic voice produced by a static female agent (note: these were 65+ aged seniors in good health, Cordasco et al. 2014), and that such seniors are less enthusiastic than impaired people in recognizing the agent's usefulness (Yaghoubzadeh et al. 2013). To our knowledge, the only comparison among user's age, definitely systematized affording a categorization of specific agent features that may engage senior and young users differently was provided by Straßmann & Krämer (2017). This study was "*a qualitative interview study with five seniors and six students*" (Straßmann & Krämer, 2017, p. 8) and showed that senior users prefer embodied human like agents over machine or animal like ones. Building upon this pilot study, the present study aims to provide a systematic investigation to assess the role of humanoid agent's appearance and as such potentially increase their acceptance among elders. In particular, the present study aims to:

- Assess elders' preferences in initiating conversations with a humanoid agent characterized by a given voice, age, eyes, face, gender, clothing style, winsomeness, and non-emotional facial expressions.
- Summon elders' responses arising from the agents' non-verbal behavior according to the pragmatic and hedonic dimensions of an interactive system firstly introduced by Hassenzahl through the *AttrakDiff* questionnaire (2014), and further en-riched in this study by new items developed by the authors (more details in Section 2.3).
- Measure elders' interest in favor of a lasting interaction with such humanoid virtual agents endowed of specific human like features.

The final aim is to test whether elders would accept a hypothetical interaction with artificial humanoid agents and, consequently, whether these agents may function as a tool for entertainment, assistance, and company.

## 2 Material and Methods

Two experiments were conducted in order to appraise the results of the present research. The first experiment was devoted to assessing physical and social agents' features among a non-senior largely age range population of subjects. This experiment was a search for consensus to substantiate the selection made by the experts. It was attempting to identify agreements (in a differently aged papulation) toward agents' characteristics such as eyes, face, voice, hair, perceived age, clothes, formal dressing, winsomeness, juvenility, look, and adjutant abilities in order to estimate substantial changes or accords in seniors' preferences. The second experiment was devoted to evaluating elders' preferences of the same agents on the hedonic and pragmatic dimensions of the interactive system.

### *2*.1 Stimuli

In order to conduct these experiments four virtual agents were defined. The virtual agents were selected from the website BOTLIBRE (www.botlibre.com) that allows users to freely create a customer service virtual agent according to their preferences and goals, providing a wide set of agents with different visual semblances. The selection of the agents was made by three experts on the basis of preferences dictated by the agents' professional and non-emotional appearance. The selected four virtual agents two males and two females, as illustrated in Figure 1, named Michael (Figure 1a), Eddie (Figure 1b), Julie3 (Figure 1c) and Victoria2 (Figure 1d) respectively, received 100% of agreements among the experts.

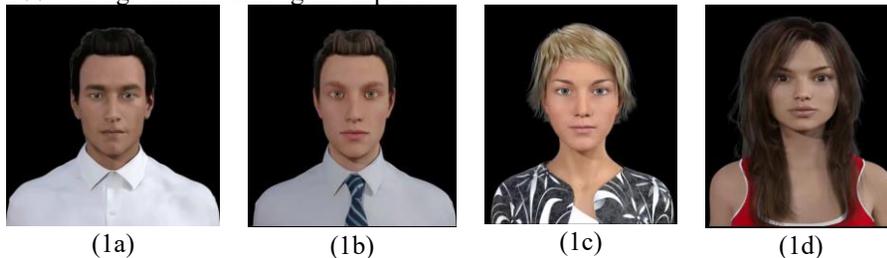

(1a)        (1b)        (1c)        (1d)

Figure 1: The four selected agents: 1a) Michael (renamed Michele); 1b) Eddie (renamed Edoardo); 1c) Julie3 (renamed Giulia); 1d) Victoria2 (renamed Clara).

The agents were depicted half torso, with definite clothing. To contextualize the agent in the local culture (the experiments were conducted in Campania a south region of Italy) they were renamed Michele, Edoardo, Giulia and Clara respectively. Each agent was provided with a different synthetic voice, producing the following Italian sentence "Hi, my name is Clara / Edoardo / Giulia / Michele. If you want, I would like to assist in your daily activity!". The synthetic voice was created through the website Natural Reader ([www.naturalreaders.com](www.naturalreaders.com)) that allows converting text to speech. The voices (recorded using the free software Audacity) were embedded into each agent's videoclip which had an average duration of about 6 seconds.

## 2.2 Participants

The first experiment involved 20 participants (13 females) aged between 27 and 64 years (mean age= 40 years, SD=±11) recruited in Salerno, a city of the Campania region, in the south of Italy. They were all in good health, with no hearing and eyesight problems (appropriately corrected with glasses in some cases) and their task was to assess physical and social features of the selected virtual agents, e.g. eyes, face, voice, hair, perceived age, clothes, formal dressing, winsomeness, juvenility, look, and adjutant abilities, on the basis of an ad hoc questionnaire created by the authors.

The second experiment involved 46 seniors (24 females), all aged 65+ years (mean age= 71.58, SD= ±6.31) recruited in Caserta a city of the Campania region, in the south of Italy. Also, in this case, participants were in good health, with no hearing and eyesight problems (appropriately corrected with glasses in some cases), and their task was to assess their preferences toward the four selected virtual agents on the pragmatic and hedonic dimensions of the interactive system (see section 2.3 for more details), their willingness to initiate a lasting interaction with them, and theirs preferred technological device to do so. Participants in both experiments accepted to join on a voluntary basis, and signed an informed consent formulated in accord with the privacy and data protection procedures established by the current Italian and European laws. The ethical committee of the Department of Psychology at the Università degli Studi della Campania, "Luigi Vanvitelli", authorized this research with the protocol number 25/2017.

## 2.3 Tools and Procedures

For the first experiment, stimuli and questionnaire were administered in the relaxation rooms of gyms and associations. Participants were asked to watch each agent's video clip and immediately after to complete a questionnaire composed of 16 items. The first part of the questionnaire consisted of 11 items and was devoted to collecting participants' opinions about the pleasantness of agents' physical and social features such as their eyes, face, voice, hair, perceived age, clothes, formal dressing, winsome-ness, juvenility, look, and adjutant abilities attributed to the agents; all rated on a 5-point Likert scale ranging from 1=not at all, to 5= very much (as an example consider the question: "Did you like the agent's eyes?"). The second part of the questionnaire was devoted to assessing the type of professions participants would endorse to the agents, among welfare, housework, protection and security, and front office jobs. Also, in this case participants' answers were rated on a 5-point Likert scale ranging from 1= I think the agent is unsuitable for this task, to 5= I think the agent is very suitable for this task.

The second experiment was devoted to assess seniors' preferences toward each of the proposed agents. The administration of stimuli was carried out either in participants' private dwellings, or day-care facilities for older people. To accomplish this goal, an ad-hoc questionnaire was developed by the authors, structured in the following clusters:

1) Cluster 1 was devoted to collecting participants' socio-demographic information, and their degree of familiarity and understanding toward smartphones, tablets, and laptops;

2) Cluster 2 collected participants' willingness to be involved in a long-lasting interaction with each of the proposed agents. This section was clustered in four sub-clusters, each consisting of 10 items, investigating the practicality, pleasure, feelings, and attractiveness experienced by participants while watching the agents' video-clips. The items proposed in each sub-cluster were inspired by Hassenzahl's theoretical model underpinning the qualities an interactive system should possess in order to receive a high acceptance from the user (2004). According to this model, a user's perception of interactive systems varies along two dimensions:

a) The system's pragmatic qualities (PQ), which focus on the usefulness, usability, and accomplishment of the tasks of the proposed system. A system receiving high scores in the PQ dimension is intended to be perceived by the user as well structured, clear, controllable, efficient, and practical.

b) The system's hedonic qualities (HQ), which focus on motivations, i.e. the reason why a user should own and use such an interactive system, (hedonic quality stimulation (HQS), identification, i.e. how captivating, as well as, of good taste the system appears, (hedonic quality identification (HQI), A system receiving high scores in the HQS and HQI sub-clusters is meant to be original, creative, captivating as well as presentable, professional, of good taste, and bringing users close to each other. These pragmatic and hedonic dimensions affect the subjective perception of the system's attractiveness (ATT) and give rise to behaviors as increased use, or dissent, as well as, emotions as happiness, engagement, or frustration. Please note, HQS will be substituted in the following with HQF (where F stands for feelings).

Cluster 2 of the proposed questionnaire is therefore organized in 4 sub-clusters, devoted to measure respectively the pragmatic quality (PQ), the hedonic identification (HQI), the hedonic feeling (HQF), and the attractiveness (ATT) of the four agents. The complete questionnaire had two more clusters respectively devoted to assessing the type of professions seniors would endorse to the agents, among which were welfare, housework, protection and security, and front office jobs, and agents' age preferences. These two last clusters were developed by the authors after the data collection had started and thus they were not administered. Future works will, however, include such data.

Each questionnaire item required a response given on a 5-point Likert scale with 1=strongly agree, 2=agree, 3=I don't know, 4=disagree, and 5=strongly disagree. Since both the second and third section of the questionnaire contained positive and negative items evaluated on a 5-point Likert scale, scores from negative items were corrected in a reverse way. This implies that low scores summon to positive evaluations, whereas high scores to negative ones. For this second experiment, participants were first asked to provide answers to the items of cluster 1, then they were asked to watch each agent's video-clip and immediately after to complete the items from cluster 2. Finally, items from cluster 3 were administered after they had seen all the four agents.

## 3 Results

This section summarizes the results of both the first and second experiments. First how naïve users assessed the physical features of the proposed agents is investigated. We will see the role played by physical and social features, as well as, adjutant abilities and professions attributed to the agents. Then, we will see how the agents have been evaluated on the pragmatic, hedonic, and attractiveness dimensions by senior users and try to correlate these preferences to the physical and social qualities of the proposed agents in order to provide explanations to the motivations and preferences offered by seniors. The scores obtained from both questionnaires were analyzed through a repeated measure ANOVA analysis, where participants' gender was considered as between factor, and agents' physical and social features as within factors. In the second experiment, participants' technology savviness was also accounted for as a between factor. Main differences among group means were assessed through Bonferroni's post hoc tests.

### 3.1 Results on agents' physical and social features, and careers (experiment 1)

The physical and social agents' features under assessment were eyes, face, voice, hair, perceived age, clothes, formal dressing, winsomeness, juvenility, look, and adjutant abilities. Participants were asked to rate how much they "enjoyed/endorsed" each of these agent's features on a 5-point Likert scale ranging from 1=not at all, to 5= very much.

It was found that agents' eyes scored significantly different, ($F(3,54)=6.215$, $p<.01$). Bonferroni post hoc tests showed that Clara's eyes (mean= 2.89) scored significantly lower ($p<.01$) than Giulia's (mean= 4.02) and Michele's eyes (mean= 3.94).

Agents' faces were significantly different ($F(3,54)=3.875$ $p<.05$). In particular, differences ($p<.05$) concerned Clara's (mean=3.33) and Michele's faces (mean=4.15), where Clara's face was rated the least enjoyable.

Also voice ($F(3,54)=6.143$ $p<.01$), showed significant differences. These differences ($p<.01$) were between Clara (mean=2.26) and Giulia (mean=3.85), with a clear preference for Giulia's voice.

The attributed age was found to be significantly different ($F(3,54)=13.279$, $p<.01$) among the agents. Clara was perceived as the younger, receiving an attributed mean age of 24.6 years, which differed significantly from Giulia (mean age =29.4), Edoardo (mean age=29.8) and Michele (mean age= 30.7).

Furthermore, agents' clothes differed significantly ($F(3,54)=6.539$, $p<.01$) because Clara's clothes (mean= 2.725) were significantly less appreciated when compared with Edoardo's (mean= 3.874, $p<.01$) and Michele's clothes (mean= 3.714, $p<.05$).

Winsomeness was found to be significantly different ($F(3,54)=6.077$, $p<.05$): the main differences emerged between Clara (mean=2.60) which was considered less winsome than Giulia (mean=3.65) and Michele (3.69).

Significant differences were found among the agents regarding them being formally dressed, ($F(3,54)=4.182$, $p<.05$). Clara was considered the least (mean=1.48) and Edoardo (mean=2.75) the most formally dressed ($p<.05$).

Participants endorsed significantly different adjutant abilities ($F(3,54)=6.899$, $p<.05$) to the agents. Clara was considered the least able to act as an assistive agent (mean=2.66) compared to Giulia (mean=3.80), Edoardo (mean=3.77) and Michele (mean= 3.82).

Finally, significant differences were found between agents' hair ($F(3,54)=3.878$, $p<.05$), look ($F(3,54)=4,728$, $p<.01$), and juvenility ($F(3,54)=3.636$, $p<.05$). However, Bonferroni post-hoc tests did not reveal any significance concerning those variables.

As for the data concerning the second part of the questionnaire, focusing on selecting which of the agents was judged more suitable to explicate either welfare, housework, protection and security, or front office jobs, it was found that Giulia (mean=3.05) was considered the most suitable to perform a housework occupation compared to Edoardo (mean=1.81) and Michele (mean=2.07), justifying the significant differences detected by the ANOVA ($F(3,54)=4.973$, $p<.01$). Michele (mean=3.78) was considered the most appropriate for protection and security ($F(3,54)=6.324$, $p<.01$), compared to Giulia (mean=3.11) and Clara (mean=2.67). Front desk occupations ($F(3,54)=4.893$, $p<.01$), were considered more appropriate for Michele (mean=3.96) compared to Clara (mean=3). Despite significant ANOVA differences with respect to welfare occupations ($F(3,54)=3.190$, $p<.05$), Bonferroni post hoc tests did not show a significant difference among agents, even though Giulia received higher mean scores (3.30).

## 3.2 Results on seniors' preferences on the pragmatic, and hedonic, dimensions of the proposed interactive systems (experiment 2)

This second experiment involved senior participants aged 65+ years being in good health. An ad-hoc questionnaire was developed with the aim to assess seniors' preferences toward the most used technological device (note: possible choices included smartphones, tablets, and laptops) and to evaluate their preferences concerning agents' pragmatic, hedonic, and attractiveness dimensions, following the definitions reported in section 2.3.

### 3.2.1 Descriptive Statistics

Analyzing the frequency of use of the proposed technological devices, it was shown that smartphones were regularly used by 56.5% of the target population, frequently but not always used by 32.6%, and never used by 10.9%. Tablets were regularly used by 4.3% of the users, frequently but not always used by 4.3%, and never used by 91.4% of the users. Laptops were regularly used by 8.7% of the target population, frequently but not always used by 30.4%, and 60.9% of the respondents had never used them. For sake of clarity this data are depicted in Figure 2 (left).

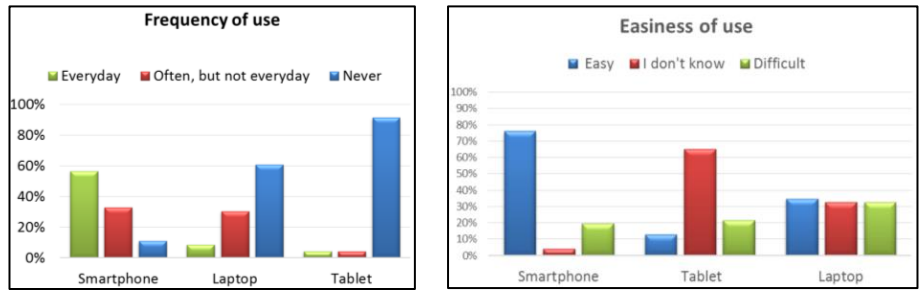

Figure 2: Descriptive Statistics: (left)Target population's frequency of use for smartphones, tablets, and laptops; (right): Easiness of use for smartphones, tablets, and laptops

Participants were also asked to evaluate their perceived easiness of using each technological device. In this context, smartphones were considered the easiest to use by 76.1% of the respondents, followed by laptops (34.8%) and tablets (13%), as depicted in Figure 2 (right).

### 3.2.2 Virtual agents' assessment

A repeated measure ANOVA analysis was conducted on the scores obtained from each questionnaire cluster and sub-cluster, considering the gender of the target population and their degree of experience with technology as between factors. The degree of experience with technology was determined separating the participant sample into two groups: participants who used a technological device often or every day were considered at "high level of expertise", while participants who used a technological device rarely or never were considered at "low level of expertise". The scores obtained by each agent on each questionnaire cluster, i.e. on the acceptance to interact with agents (cluster1), on the pragmatic (PQ), hedonic quality identity (HQI), hedonic quality feelings (HQF), and attractiveness (ATT) dimensions were considered as within factors. The significance was set at $\alpha= .05$. The scores obtained from the questionnaire's negative items were reversed so that lower scores indicate strong preferences and higher scores low preferences toward the agent dimensions defined by the questionnaire clusters.

**Willingness to interact with the agents**

Significant differences were found among agents ($F(3,126)=16.323$, $p<<.001$) on the seniors' willingness to interact with them. Bonferroni post hoc tests revealed that these differences were due to significant differences in the scores obtained by Giulia (mean score=1.40) with respect to Clara (mean score=1.85, $p<.05$), Edoardo (mean score=2.40, $p<<.001$) and Michele (mean score=2.58, $p<<.001$), as well as, significant different scores obtained by Clara compared to Edoardo ($p<.05$) and Michele ($p<.05$). No significant differences were found between Edoardo and Michele.

An interaction was found between the participants' gender and their willingness to interact with the virtual agents ($F(3,126)=4.323$, $p<.05$). Bonferroni post hoc tests revealed this was because male and female participants differed significantly in their willingness to interact with Michele ($p<.05$), being females less available (mean score=3.08) than male participants (mean score=2.09) to interact with them.

**Pragmatic Qualities (PQ)**

The pragmatic qualities seniors attributed to the agents differed significantly ($F(3,126)=24.530$, $p<<.001$). Bonferroni post hoc tests attributed these differences to significant differences in the scores obtained by Giulia (mean score=18.84) in comparison to Edoardo (mean score=32.52, $p<<.001$), and Michele (mean score=33.36, $p<<.001$), as well as, significant different scores obtained by Clara (mean score= 22.23) with respect to Edoardo ($p<.05$) and Michele ($p<<.001$). No significant differences were found between Giulia and Clara and between Edoardo and Michele.

**Hedonic Qualities – Identity - (HQI)**

Agents' HQI scores were significantly different ($F(3,126) =22.5$, $p<<.001$). Bonferroni post hoc tests revealed that these differences were caused by significant differ-ences in the scores obtained by Giulia (mean score=19.97) compared to Edoardo (mean score=34.62, p<<=.001) and Michele (mean score= 33.67, p<<.001), as well as, Clara (mean score= 24.38) compared to Edoardo (p<.05) and Michele (p<.05). No significant differences were found between Giulia and Clara and between Edoardo and Michele. An interaction emerged between participants' gender and their degree of experience with technology (F(1,42)=4.687, p< .05). Bonferroni post hoc tests revealed that female participants with high technological experience (mean=30.35) attributed to the agents higher HQI scores (p<.05) (and therefore were less prone to identify themselves with the agents) than female participants with low technological experience (mean score=26.66).

### Hedonic Qualities – Feelings - (HQF)

Agents' HQF scores differed significantly (F(3,126) =20.99, p<<.001). Bonferroni post hoc tests revealed that these differences were due to significantly different scores obtained by Giulia (mean score=19.41) with respect to Edoardo (mean score=33.00, p<<= .001) and Michele (mean score=32.31, p<<.001), as well as, the score obtained by Clara (mean score=23.60) with respect to Edoardo (p<.05), and Michele (p<.05). No significant differences were found between Giulia and Clara or between Edoardo and Michele.

Participants' experience with technology (F(1,42)=7.368, p<.05) also played a significant role in the assessment of the HQFs features attributed by seniors to the agents. Bonferroni post hoc tests revealed significant differences between seniors highly technological experienced with respect to low experienced ones (p<.05), the first group scored the agents (mean=28.47) less exciting, appealing, and captivating than the second one (mean score=25.69).

### Attractiveness (ATT)

Attractiveness was also found to be significantly different (F(3,126)=25.127, p<<.001) among the agents. Bonferroni post hoc tests revealed that these differences were caused by better scores obtained by Giulia (mean score=19.14) in comparison with Edoardo (mean score=33.91, p<<.001) and Michele (mean score=32.62, p<<.001), as well as, by Clara (mean score=22.49) in comparison with Edoardo (p<<.001) and Michele (p<<.001). No significant differences were found between Giulia and Clara or between Edoardo and Michele.

Participants' experience with technology was also significantly different (F(1,42) =4.288, p<.05). Bonferroni post hoc tests revealed that participants with high technological experience (mean=28.15) scored the agents less attractive than those with low technological experience (mean score=25.93). In addition, a significant interaction was found between participants' gender and their level of experience with technology (F(1,42)=4.192, p<.05), because (Bonferroni post hoc tests) female participants with high technological experience find the agents less attractive (mean=29.31) than female participants with low technological experience (mean score=24.89, p<.05).

## 4 Discussions and Conclusion

The data of the first experiment showed that specific human like features such as eyes, face, voice, perceived age, clothes, formal dressing, adjutant abilities, and winsomeness affect user's preferences toward agents, while hair, juvenility, and look do not seem to play any role on user's acceptance. Users also expressed preferences on the type of job the proposed agents are most suitable at, suggesting that they attributed to agent's specific competences and capabilities. From these data, it has been learned that young looking agents (such as Clara) are evaluated as ingenuous, and inexperienced and therefore may not serve to specific purposes or occupy specific positions. However, this first experiment served just as a means to assess agents' physical and social features over a large age range of the target population.

The second experiment was instead devoted to assessing elders' preferences, involvement, and engagements with such agents. The interesting results were that no matter how inappropriate, young, and inexperienced, Clara was considered by the non-elder population of the first experiment, seniors clearly expressed their willingness to interact with her rather than with the two male agents. This was true on both the pragmatic (seniors considered Clara better designed, more unmistakable, controllable, user-friendly, and efficient than her male counterparts) and hedonic (seniors judged Clara more captivating, exciting, engaging than her male counterparts) dimensions.

Our data underlined a strong preference of seniors toward female humanoid agents, independently from their gender and their technology savviness. Between the two proposed female agents, seniors' preferences toward Giulia scored statistically significantly higher than those attributed to Clara. It can be hypothesized that Giulia is preferred to Clara because Clara may have been perceived younger, more ingenuous, and less professional as from the results of the first experiment. Giulia is always rated significantly better than Clara, Michele and Edoardo. This is true for the pragmatic qualities (PQ), i.e. for facets inherently concerning the practicality, expertise and usefulness of the agent; and the hedonic qualities (HQI and HQF), i.e. for features regarding the individual identity of the agent as well as the feelings that can emerge during a potential interaction with the agent. Lastly, the female agents seem more engaging than male ones on the attractiveness (ATT) dimension.

The above reported data suggests as conclusion of this study, that seniors' willingness to be assisted by a virtual agent is strongly affected by the gender of the proposed agent. Starting from this data we tried to figure out why our seniors behaved like that. Seniors are more willing to interact and be assisted by female agents. We might argue that this maybe the results of social and cultural wisdoms with respect to the individual perception of others and the role females occupy in society, intended here as the social experience, culture, and environment in which an individual grows. These elements provide the first cognitive and learning understandings about others and therefore, about the social role of men and women and their social differentiation in the community. Gender is the way in which historically and socially in a given community, a (mutable) meaning is attributed to the physical and biological differences between human beings. These differences are converted in artefacts of human activities by the cultural environment, establishing a form of labour division, initiated by the prehistoric gathering-hunter division, and exemplified by modern complex societies, through cultural distinctions among masculine and feminine professions (Signorelli, 2011). Therefore, because of their specific cultural heritage and the role of women in the specific society of the target population, it is expected that seniors may prefer a female assistive virtual adjutant. Change in the cultural heritage and social rules may however alter such preferences.

The proposed work attempts to assess elders' preferences in initiating conversations with humanoid agents and provide measures of their interest in favor of a long-lasting interaction, as well as, their degree of acceptance, and their willingness to exploit these agents as tools for entertainment, assistance, and company. To this aims it provides original data suggesting a seniors' preference toward female agents, and a disenchantment of the most technologically experienced ones on the agents' hedonic and attractiveness qualities. Seniors' savviness of technological devices is an impediment to feel captivated, engaged, and rise behaviors of increased use, or dissent, as well as, positive or negative emotions**.** The research also offers to the international scientific community an original assessment tool (the questionnaire developed by the authors) that deliver a systematic way for conducting such investigations, in order to identify further differences caused by social, cultural, and environmental factors. This is the way this research was envisaged, since it was conducted as part of an EU funded project, underlining the limitations associated to the present results. Currently, the preferences identified in this study refer to the South Italian population located in the Campania region. We do not yet know whether these results may be different for the Northern Italian senior population, neither do we know about potential differences found in other EU countries. The research undertaken by the Empathic project (https://www.cordis.europa.eu/project/rcn/212371_en.html) is intended to answer some of these questions, and consequently offer insights that can support the foundation of reliable predictions on seniors' acceptance of virtual humanoid interfaces for assistive uses.

**ACKNOWLEDGMENTS**

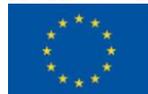 The research leading to the results presented in this paper has been conducted in the project EMPATHIC (Grant N: 769872) that received funding from the European Union's Horizon 2020 research and innovation programme.

14	Anna Esposito et al.